# Recover plaintext attack to block ciphers

Li An-Ping

Beijing 100085, P.R.China
apli0001@sina.com

**Abstract**

In this paper, we will present an estimation for the upper-bound of the amount of 16-bytes plaintexts for English texts, that is not sufficient large make clear that the block ciphers with block length no more than 16-bytes will be subject to recover plaintext attacks in the occasions of plaintext -known or plaintext-chosen attacks.

# 1. Introduction

For the security of block ciphers there are many researches, which may be found in most of textbooks and papers in cryptography, refer to see [1]. It is known that block ciphers have a characteristic that it encrypt plaintexts in blocks with a regular encryption scheme, so, plaintexts are 1-1 related to the ciphertexts in blocks for a secret key. It is not difficult to know that these block ciphers will be easy subjected to recover plaintext attack if the amount of block plaintexts is not sufficient large. Suppose that the amount of all possible plaintexts blocks is no more than $2^m$, an adversary has a dictionary of the block-pairs (ciphertext, plaintext) with size about $2^{m/2+2}$, then he will recover a block plaintext while collect $2^{m/2}$ blocks of ciphertexts with high successful probability by the general birthday paradox.

In most of the currently used block ciphers, the output sizes, that is, the lengths of blocks are equal to, or smaller than 16 bytes. In this paper, we will show that in the case of English text the number of 16-bytes plaintexts is less than $2^{56}$, so the block ciphers with output size of 16 bytes will be vulnerable to recover plaintext attacks in the occasions of plaintext-known or plaintext-chosen attacks.

In the rest of this section, we give some conceptions used in this paper.

Denoted by $\mathcal{Q}$ the vocabulary for the plaintexts, and suppose that the size $|\mathcal{Q}| = N$. For a word $w \in \mathcal{Q}$, , denote by $|w|$ the length, i.e., the number of the letters contained in the word $w$.

An English phase or a plaintext block $\alpha$ is called of $k$-terms if it consists of $k$ words or parts of words, There are four possible expressions for the $k$-terms plaintext blocks

$$word_1 \vee word_2 \vee \cdots \vee word_k \qquad (1.1)$$
$$word_1 \vee word_2 \vee \cdots \vee word_k \vee \qquad (1.2)$$
$$\vee word_1 \vee word_2 \vee \cdots \vee word_k \qquad (1.3)$$
$$\vee word_1 \vee word_2 \vee \cdots \vee word_k \vee \qquad (1.4)$$

Where $word_i$ is the $i$ th word of $\alpha$, and symbol $\vee$ represents a blank space.

It should be mentioned that there are the possibilities that $word_1$ in (1.1), (1.2) and $word_k$ in (1.1), (1.3) are not complete English words but only parts. Besides, possibly there are existed some blocks contain some punctuation marks such as ',' or '.' or ';', which will be agreed to be a character rather than a term, except the special case that $word_1$ in (1.1) or (1.2) is just a punctuation mark. We will only take the frequently used three punctuation marks ',' , '.' and ';' into the consideration in the following discussion.

For the simplicity, in this paper, it is assumed that the words in the vocabulary $\mathcal{Q}$ are consist of English letters, no include special characters such as @, #, etc, and Arabian numbers and abbreviations.

# 2. The estimations of the amount of plaintext blocks

In this section, we will present a estimation for the amount of 16-bytes plaintexts.

**Proposition 1.** Suppose that $\mathcal{Q}$ is a vocabulary consist of English words, including no special characters and Arabian numbers, and the size $|\mathcal{Q}| \leq 60000$. Let $\mathcal{F}$ be the set of all possible 16-bytes blocks of English texts over $\mathcal{Q}$. Denoted by $\mathcal{Q}_i$ the subset of $\mathcal{Q}$ consist of $i$-letters words, $1 \leq i \leq 16$, if the distribution of $|\mathcal{Q}_i|$ satisfy that

$$|\mathcal{Q}_i| \leq \mu \cdot C_{16}^{i-1}, \quad 1 \leq i \leq 16, \tag{2.1}$$

where $\mu$ is a constant, $\mu = 2$, then

$$|\mathcal{F}| \leq 2^{56}. \tag{2.2}$$

Proof. Denoted by $\tilde{\mathcal{F}}$, $\dot{\mathcal{F}}$ and $\mathcal{F}'$ the subsets of $\mathcal{F}$ consist of 16-bytes plaintexts with that the first letter is a minuscule one, a capital one and a punctuation respectively. We will see that $\tilde{\mathcal{F}}$ possess a main part in the amount. For an positive integer $k$, $1 \leq k \leq 8$, Let $\mathcal{F}_k$ be the subsets of $\tilde{\mathcal{F}}$ consist of $i$-terms blocks, and $\mathcal{F}_k^{(1)}$, $\mathcal{F}_k^{(2)}$, $\mathcal{F}_k^{(3)}$ and $\mathcal{F}_k^{(4)}$ be the subsets of $\mathcal{F}_k$ with the expression forms (1.1), (1.2), (1.3) and (1.4) respectively. We will firstly calculate $|\mathcal{F}_k^{(1)}|$.

Suppose that $\zeta \in \mathcal{F}$, is a $k$-terms block,

$$\zeta = (\vee) word_1 \vee word_2 \vee \cdots \vee word_k (\vee). \tag{2.3}$$

Denoted by $|word_i| = c_i$, obviously, $c_i \geq 1$, $1 \leq i \leq k$, and

$$\sum_{1 \leq i \leq k} c_i = 16 - k + \delta, \tag{2.4}$$

$\delta = 1, 0, 0, -1$, for $\zeta \in \mathcal{F}_k^{(1)}, \mathcal{F}_k^{(2)}, \mathcal{F}_k^{(3)}, \mathcal{F}_k^{(4)}$ respectively. At first we are restricted in the case $\zeta \in \mathcal{F}_k^{(1)}$. We call $\zeta$ is of $(c_1, c_2, ..., c_k)$-type, and let $\mathcal{F}_{(c_1, c_2, \cdots, c_k)}$ be the set of $(c_1, c_2, ..., c_k)$-type blocks. For any $i$-subset $I$ of $\{1, 2, \cdots, k\}$, denoted by $\mathcal{F}_{(c_1, c_2, \cdots, c_k)}^{(I)}$ be the subset of $\mathcal{F}_{(c_1, c_2, \cdots, c_k)}$ consist of the blocks with $i$ punctuation marks following the words with indices in $I$. In the next, we calculate the sizes $|\mathcal{F}_{(c_1, c_2, \cdots, c_k)}^{(i)}|$, and we at first calculate $|\mathcal{F}_{(c_1, c_2, \cdots, c_k)}^{(0)}|$.

Denoted by $x = \min(26^{c_1}, N)$, $y = \min(26^{c_k}, N)$, by (2.1), it has

$$|\mathcal{F}^{(0)}_{(c_1,c_2,\cdots,c_k)}| \le x \times y \times \prod_{i=2}^{k-1}(\mu \cdot C_{16}^{c_i-1})$$

$$\le (x/C_{16}^{c_1-1}) \times (y/C_{16}^{c_k-1}) \times \mu^{k-2} \times \prod_{i=1}^{k} C_{16}^{c_i-1} . \tag{2.5}$$

By the basic combinatorics, we know that for any positive integer $s$, it has

$$\sum_{c_1+\cdots+c_k=s} \prod_{i=1}^{k} C_{16}^{c_i} = C_{16k}^{s} . \tag{2.6}$$

And for the assumption $N \le 60000$, it is easy to know that

$$\min\{26^{c_1}/C_{16}^{c_1-1}, N/C_{16}^{c_1-1}\} \le 26^3/C_{16}^{2} \le 147, \tag{2.7}$$
$$\min\{26^{c_k}/C_{16}^{c_k-1}, N/C_{16}^{c_k-1}\} \le 26^3/C_{16}^{2} \le 147.$$

So, with (2.5), (2.6), (2.7) and (2.4), we have

$$\sum_{c_1+\cdots+c_k=17-k} |\mathcal{F}^{(0)}_{(c_1,c_2,\cdots,c_k)}| \le (147)^2 \times (\mu)^{k-2} \times C_{16k}^{17-2k} \tag{2.8}$$

To get an estimation for $\mathcal{F}^{(I)}_{(c_1,c_2,\cdots,c_k)}$, $i > 0$, have to change $16$ into $16-i$ in the equation (2.4), and notice that there are $C_k^i \cdot 3^i$ $i$-subsets $I$'s, so we have

$$\sum_{I} \sum_{c_1+\cdots+c_k=17-i-k} |\mathcal{F}^{(I)}_{(c_1,c_2,\cdots,c_k)}| \le (147)^2 \times C_k^i \times 3^i \times \mu^{k-2} \times C_{16k}^{17-i-2k} \tag{2.9}$$

Hence,

$$|\mathcal{F}_k^{(1)}| \le \sum_{0 \le i \le k} (147)^2 \times C_k^i \times 3^i \times \mu^{k-2} \times C_{16k}^{17-i-2k}$$

$$\le (147/\mu)^2 \times \sum_{0 \le i \le k} C_k^i \times 3^i \times \mu^k \times \frac{(16k)^{17-2k}}{(17-2k)!} \times \frac{(17-2k)\cdots(17-i-2k+1)}{(16k)^i}$$

$$\le \frac{(147/\mu)^2 \cdot \mu^k}{\sqrt{2\pi(17-2k)}} \times \left(\frac{e \cdot 16k}{17-2k}\right)^{17-2k} \times \sum_{0 \le i \le k} C_k^i \times 3^i \times \frac{(17-2k)\cdots(17-i-2k+1)}{(16k)^i} \tag{2.10}$$

$$\le \frac{(147/\mu)^2 \cdot \mu^k}{\sqrt{2\pi(17-2k)}} \cdot \left(\frac{e \cdot 16k}{17-2k}\right)^{17-2k} \cdot \left(1 + \frac{3 \cdot (17-2k)}{16k}\right)^k ,$$

where Stirling's formula has been applied.

The estimations for $|\mathcal{F}_k^{(2)}|$, $|\mathcal{F}_k^{(3)}|$ and $|\mathcal{F}_k^{(4)}|$ are similar to the one above, but notice that that $word_1$ in (1,2), (1.4) and $word_k$ in (1.3), (1.4) are complete English word rather than a part, and now $\delta = 0$ in the equation (2.5) for $|\mathcal{F}_k^{(2)}|, |\mathcal{F}_k^{(3)}|$ and $\delta = -1$ for $|\mathcal{F}_k^{(4)}|$. Thus, we has

$$|\mathcal{F}_k^{(i)}| \leq \frac{(147/\mu)\cdot\mu^k}{\sqrt{2\pi(16-2k)}} \cdot \left(\frac{e\cdot 16k}{16-2k}\right)^{16-2k} \cdot \left(1+\frac{3\cdot(16-2k)}{16k}\right)^k, \qquad i=2,3.$$

$$|\mathcal{F}_k^{(4)}| \leq \frac{\mu^k}{\sqrt{2\pi(15-2k)}} \cdot \left(\frac{e\cdot 16k}{15-2k}\right)^{15-2k} \cdot \left(1+\frac{3\cdot(15-2k)}{16k}\right)^k.$$

(2.11)

So,

$$\begin{aligned}
|\tilde{\mathcal{F}}| &\leq \sum_{1\leq k\leq 8}(|\mathcal{F}_k^{(1)}|+|\mathcal{F}_k^{(2)}|+|\mathcal{F}_k^{(3)}|+|\mathcal{F}_k^{(4)}|) \\
&\leq \sum_{1\leq k\leq 8}\frac{(147/\mu)^2\cdot\mu^k}{\sqrt{2\pi(17-2k)}}\cdot\left(\frac{e\cdot 16k}{17-2k}\right)^{17-2k}\cdot\left(1+\frac{3\cdot(17-2k)}{16k}\right)^k \\
&+ \sum_{1\leq k< 8}2\times\frac{(147/\mu)\cdot\mu^k}{\sqrt{2\pi(16-2k)}}\cdot\left(\frac{e\cdot 16k}{16-2k}\right)^{16-2k}\cdot\left(1+\frac{3\cdot(16-2k)}{16k}\right)^k \\
&+ \sum_{1\leq k< 8}\frac{\mu^k}{\sqrt{2\pi(15-2k)}}\cdot\left(\frac{e\cdot 16k}{15-2k}\right)^{15-2k}\cdot\left(1+\frac{3\cdot(15-2k)}{16k}\right)^k + (147)^2\times\mu^6
\end{aligned}$$

(2.12)

For $\mu=2$, it has

$$|\tilde{\mathcal{F}}|\leq 3.73\times 10^{16}.$$

(2.13)

In respect to the estimation of $|\dot{\mathcal{F}}|$, we know that the first letter of a sentence is capital or minuscule is determined by the punctuation ahead it, and that the first letter of $word_1$ in (1.1) or (1.2) is capital one means $word_1$ is a complete English word rather than a part. Denoted by $\mathcal{F}^{(i)}$, $1\leq i\leq 4$, the subsets of $\tilde{\mathcal{F}}$ consist of the plaintext blocks with type as in (1.1), (1.2), (1.3) and (1.4) respectively, that is, $\mathcal{F}^{(i)}=\bigcup_k \mathcal{F}_k^{(i)}, i=1,2,3,4$. Then, it has

$$|\dot{\mathcal{F}}|\leq \frac{\mu}{147}|\tilde{\mathcal{F}}|+|\mathcal{F}^{(3)}|+|\mathcal{F}^{(4)}|\leq 4.2\times 10^{14}.$$

(2.14)

For the estimation of $|\mathcal{F}'|$, provided to substitute $16$ by $16-1$ in the estimation of $|\mathcal{F}^{(3)}|$ and $|\mathcal{F}^{(4)}|$, and with multiple $3$ for there are three punctuation marks. So

$$|\dot{\mathcal{F}}|\leq 4\times 10^{13}$$

(2.15)

Hence, we have

$$|\mathcal{F}|\leq |\tilde{\mathcal{F}}|+|\dot{\mathcal{F}}|+|\mathcal{F}'|\leq 3.8\times 10^{16}\leq 2^{56}. \qquad \square$$

*Remark 1.* It is likely the inequation (2.1) is true for the distribution of English words, but we have not checked in total, so we have taken it as a condition, so that the constant $\mu$ may be modified according to the actual cases.

*Remark* 2. Moreover, for a $k$-letters word $w$, and a positive integer $i, i \leq k$, we call the segment formed by the first $i$ letters of $w$ as the $i$-prefix of $w$, similarly, the segment formed by the last $i$ letters of $w$ as the $i$-suffix of $w$. Denoted by $\mathcal{Q}^{[i]}$ and $\mathcal{Q}_{[i]}$ the sets of all the distinct $i$-prefix's and $i$-suffix's of the words in $\mathcal{Q}$ respectively. Suppose that

$$|\mathcal{Q}^{[i]}| = \lambda^{(i)} \times C_{16}^{i-1}, \quad |\mathcal{Q}_{[i]}| = \lambda_i \times C_{16}^{i-1}, \qquad (2.16)$$

and denoted by $\lambda = \max_i \{\lambda_i, \lambda^{(i)}\}$. In the proof above we known that $\lambda \leq 147$, however, we guess that for the ordinary vocabulary there may be $\lambda = 26$, if so, then

$$|\mathcal{F}| \leq 1.8 \times 10^{15} \leq 2^{51}. \qquad (2.17)$$

It is easy to know that the conjecture is true for $i = 1$, and $i > 5$, so the rest to be verified are the cases $2 \leq i \leq 5$.

**3. Conclusion**

For the simplicity of discussion, we have excluded Arabian numbers and some special characters such as @, $, etc, and some special punctuations such as '!', '?', etc, though they occasionally appear in the English texts, but a little. So, the estimation above may be viewed as the one for the frequently appeared ones. The calculations in the paper is nearly in combinatorics, no considerations on the English grammar, logic and semantics, so it is very likely that the actual amount of plaintext blocks will be much smaller then the one presented in Proposition 1. In fact, our first idea is from the consideration in English grammar, but which is somewhat trifling.

The result presented indicate that the block ciphers with 16-bytes block length such as AES will be subject to recover plaintext attacks when applied to encrypt English texts in the occasions of plaintext-known or plaintext-chosen attacks.

From the discussion above, we have seen that the amount of plaintext blocks not only depend the size of block length but also the distribution of the words in languages.